\documentstyle[procl]{article}

\input{psfig}

\def\Journal#1#2#3#4{{#1} {\bf #2}, #3 (#4)}

\def\PLB{{\em Phys. Lett.}  B}

\def\be{\begin{equation}}

\def\ee{\end{equation}}

\def\bea{\begin{eqnarray}}

\def\eea{\end{eqnarray}}

\begin{document}
\hfill$\vtop{   \hbox{\normalsize TTP96-30}
                \hbox{\normalsize MADPH-96-955}
                   }$ 

\vspace*{5mm}
\title{NLO Corrections to Jet Cross Sections in DIS \footnote{
Talk presented by E.~Mirkes at the 'International Workshop
on Deep Inelastic Scattering and Related Phenomena' (DIS96), April 1996,
 Rome.}
}

\author{ E. MIRKES }

\address{Inst. f. Theor. Teilchenphysik, 
         Universit\"at Karlsruhe, D-76128 Karlsruhe, Germany\\}

\author{ D. ZEPPENFELD }

\address{Department of Physics, 
        University of Wisconsin, Madison, WI 53706, USA\\}





\thispagestyle{empty}
 
\maketitle\abstracts{
Next-to-leading order corrections to
jet cross sections in deep inelastic scattering at HERA are
studied. The predicted jet rates allow for a precise
determination of $\alpha_s(\mu_R)$ and the gluon density
at HERA. We argue, that the ``natural'' renormalization and
factorization scale is in general set by the average  $k_T^B$
of the jets in the Breit frame. Some implications
for the associated forward jet production in the low $x$ 
regime at HERA are  discussed.
}  
\section{Introduction}
One of the main topics to be studied at HERA is the deep
inelastic  production of multi jet 
events,
where good event statistics are expected 
allowing for precision tests of QCD~\cite{tref,rose,flam}.  
Clearly, next-to-leading order (NLO) QCD corrections
are mandatory on the theoretical side for such tests.
Full NLO corrections for one and two-jet production cross sections
and distributions are now available \cite{mz1}
and implemented in the fully differential
$ep \rightarrow n$ jets event generator MEPJET,
which allows to analyze 
arbitrary jet definition schemes and 
general cuts in terms of parton 4-momenta. 
A variety of topics can be studied with these tools. 
They include:  
a) The determination of  $\alpha_s(\mu_R)$ over a range of scales $\mu_R$
from dijet production.
b) The measurement of the 
gluon density in the proton (via $\gamma g\to q\bar q$)
c) The measurement of the 
polarized gluon density $\Delta{g}$  with polarized beams of electrons
and protons at HERA \cite{fkm}.
d) Associated forward jet production in the low $x$ 
regime as a signal of BFKL dynamics.

The importance of higher order corrections
and recombination scheme dependencies 
of the two jet cross sections
for  four different jet algorithms 
(cone, $k_T$, JADE, W) was already discussed
in Refs.~[4,6,7].
While the higher order corrections and recombination scheme
dependencies in the cone
and $k_T$ schemes are small, very large corrections can 
appear in the $W$-scheme. 
Depending on the definition of the jet resolution mass
and on the recombination scheme, the NLO cross sections in the
$W$-scheme can differ  by almost a factor of two~\cite{mz1,mz2}.
Trefzger~\cite{tref} and Rosenbauer~\cite{rose}
find similarly large differences in the experimental
jet cross sections (which are in good agreement with the theoretical
predictions from MEPJET), if the data are processed with 
exactly the same jet resolution mass
and recombination prescription as used in the theoretical calculation.
Note however, that 
the dependence on the recombination scheme first appears in the NLO 
calculation, where a jet may be composed of two partons. This internal jet
structure is thus  only simulated at tree level and thus the dependence 
of the cross section on the recombination scheme is subject to potentially
large higher order corrections.
An alternative fully differential NLO MonteCarlo program
DISENT is under construction~\cite{catani}.

Previous programs \cite{projet,disjet} were limited to a JADE type 
algorithm and are not flexible enough to implement the various
recombination schemes. 
In addition, approximations were made 
to the matrix elements in these programs which are not valid in large regions
of phase space~\cite{mz1}.
It is therefore not surprising, that one finds
inconsistent values for $\alpha_s$ using these programs \cite{rose}, 
if data are processed with different recombination schemes.

We conclude from these studies that the
cone and $k_T$ schemes appear better suited for precision
QCD tests  and will concentrate only on those schemes
in the following.

In the cone algorithm (which is defined in the laboratory frame) the distance 
$\Delta R=\sqrt{(\Delta\eta)^2+(\Delta\phi)^2}$ between two partons 
decides whether they should be recombined to a single jet. Here the variables 
are the pseudo-rapidity $\eta$ and the azimuthal angle $\phi$. We 
recombine partons with $\Delta R<1$.
Furthermore, a cut on the jet transverse momenta of $p_T(j)>5$~GeV in the lab 
frame is imposed.

For the $k_T$ algorithm (which is implemented in the Breit frame), 
we follow the description introduced
in Ref.~[11]. The hard scattering scale $E_T^2$ is
fixed to 40 GeV$^2$ and $y_{cut}=1$ is the resolution parameter 
for resolving the macro-jets.

\section{The determination of  $\alpha_s(\mu_R)$ from dijet production}
The dijet cross section
is proportional to $\alpha_s(\mu_R)$ at leading order (LO), thus suggesting 
a direct measurement of the strong coupling constant. However, the
LO calculation leaves the renormalization scale $\mu_R$ undetermined.
The NLO corrections substantially reduce the renormalization and 
factorization scale dependencies which are present in the LO calculations 
(see below)
and thus reliable cross section predictions in terms of $\alpha_s(m_Z)$
are made possible.
Clearly, a careful study of the choice
of scale in the dijet cross section
is needed in order to extract a reliable value for 
$\alpha_s(M_z)$.
Jet production in DIS is a multi-scale problem
and it is not a priori clear at which scale $\alpha_s$ is probed.
However, it was argued \cite{mz3}, that the ``natural'' scale
for jet production in DIS
is set by the average $k_T^B$
of the jets in the Breit frame.
Here, $(k_T^{B}(j))^2$ is defined by 
$2\,E_j^2(1-\cos\theta_{jP})$, where the subscripts $j$ and $P$
denote the jet and proton, respectively (all quantities are defined
in the Breit frame).
It can be shown \cite{mz3}
that $\sum_j \,k_T^B(j)$ smoothly interpolates 
between the correct limiting
scale choices, it approaches $Q$ in the parton limit and it corresponds to 
the jet
transverse momentum $p_T^B$ (with respect to the $\gamma^*$-proton direction)
when the photon virtuality becomes negligible.
It therefore 
appears to be the ``natural'' scale for multi jet production in DIS.

Fig.~1a shows the scale dependence of the dijet cross section in
 LO and NLO  for the $k_T$ scheme.
We have considered scales related to the scalar sum of the partons 
$k_T^B$ (solid curves), the scalar sum of the partons $p_T$
with respect to the $\gamma^*$-boson direction (dashed curves)
and the virtuality $Q$ of the incident photon.
The LO (NLO) results are based on 
the LO (NLO) parton distributions from GRV \cite{grv} together with
the one-loop (two-loop) formula 
for the strong coupling constant.
Kinematical cuts are imposed to closely model the H1 event 
selection. More specifically, we require 
an energy cut of $E(e^\prime)>10$~GeV on the scattered 
electron, and a cut on the pseudo-rapidity $\eta=-\ln\tan(\theta/2)$
of the scattered lepton depending on $Q^2$, i.e.
$ -2.794 < \eta(l^\prime)<-1.735$\, for $Q^2 < 100$ GeV$^2$
$ -1.317 < \eta(l^\prime)<2.436$ \,for $Q^2 > 100$ GeV$^2$.
In addition, we require 
$ -1.154 < \eta(j)< 2.436$.
The scale dependence of the dijet cross section
does not markedly improve in NLO for $\mu^2=\xi Q^2$
(dotted lines in Fig.~1a).
The resulting $\xi$ dependence for
$ \mu_R^2 = \mu_F^2 = \xi\;(\sum_i \,k_T^B(i))^2$
($ \mu_R^2 = \mu_F^2 = \xi\;(\sum_i \,p_T^B(i))^2$)
is shown  as the solid (dashed) lines in Fig.~1a.
The uncertainty from the variation
of the 
renormalization and factorization scale
scale is markedly reduced compared to 
the LO predictions.
Hence, the theoretical uncertainties due to the scale variation
are small suggesting a precise
determination of $\alpha_s(<k_T^B>)$ for different $<k_T^B>$ bins,
where 
$
<k_T^B>=\frac{1}{2}\,\, (\sum_{j=1,2} \,k_T^B(j)).
$
It has also been shown~\cite{mz3}, that the  $<k_T>$, $<p_T>$ and $Q$ 
distributions can differ  substancially 
for different $<k_T^B>$ bins (in particular for low $Q^2$),
which explains the large differences in the scale dependence in Fig.~1.
Fig.~1b shows the scale dependence of the dijet cross section in
cone scheme for  cuts similar to the  ZEUS selection:
40~GeV$^2<Q^2<2500$ GeV$^2$,
$0.04 < y < 1$,
and a cut on the pseudo-rapidity $\eta=-\ln\tan(\theta/2)$
of the scattered lepton and jets of $|\eta|<3.5$. 
The results are very similar to the results for the $k_T$ scheme
shown in Fig.~1a.
\begin{figure}[htb]
\psfig{figure=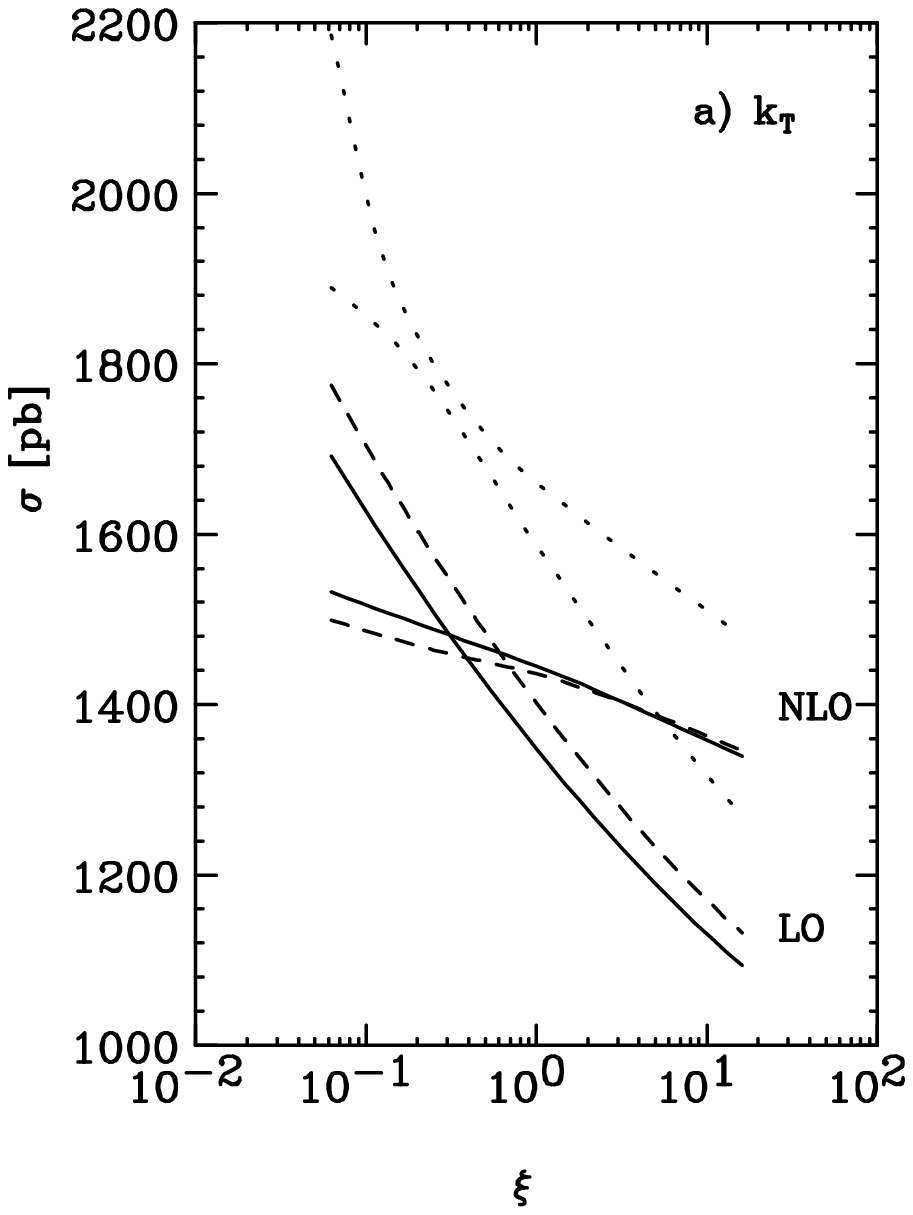,height=2.5in}
\vspace{-6.3cm}
\hspace*{5.cm}
\psfig{figure=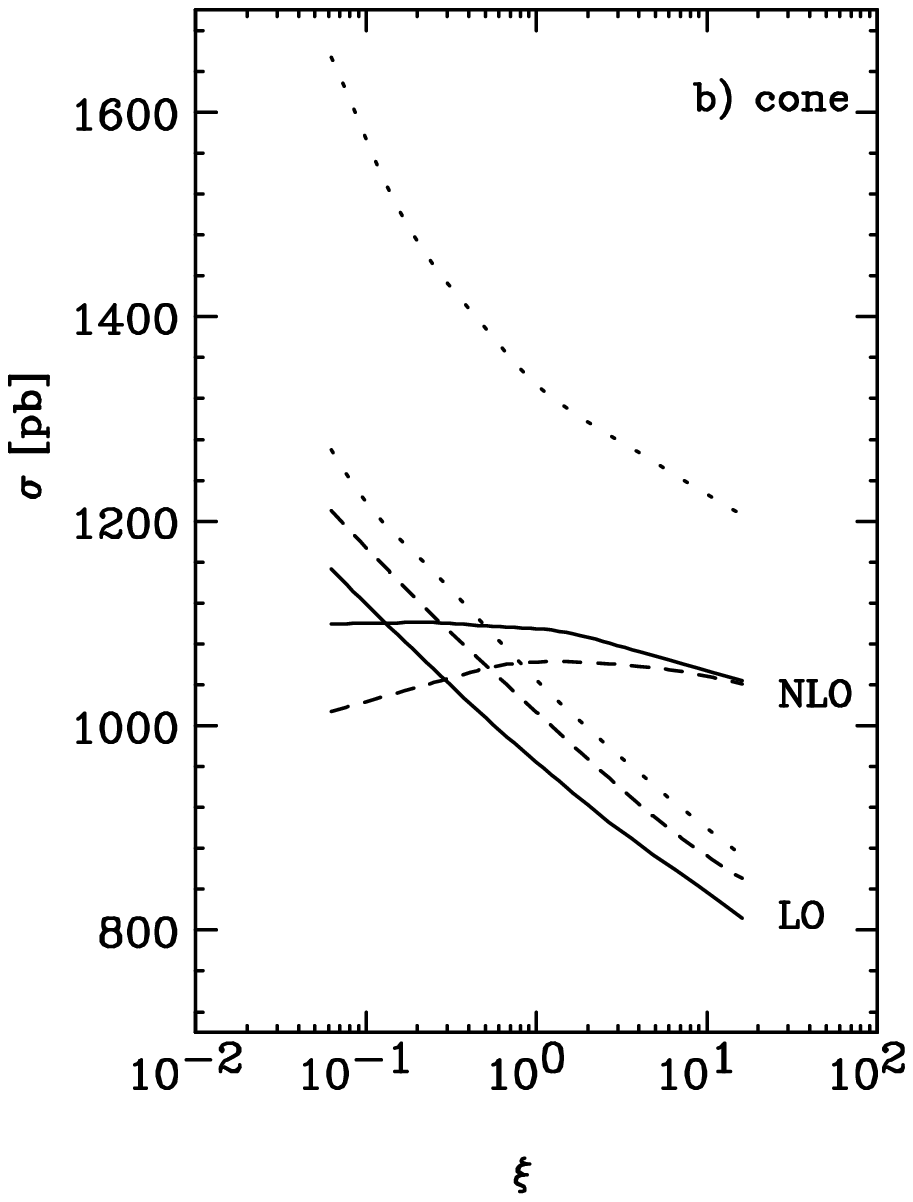,height=2.5in}
\caption{
a) Dependence of the two-jet exclusive cross 
section in the $k_T$ scheme 
on the  scale factor $\xi$.
The dashed curves are for $\mu_R^2=\mu_F^2=\xi\;(\sum_i\;p_T^B(i))^2$.
Choosing $(\sum_i\;k_T^B(i))^2$ as the basic scale yields the solid curves.
Choosing $Q^2$ as the basic scale yields the dotted curves.
Results are shown 
for the LO (lower curves) and NLO calculations; b) the same as  a) for
the cone scheme. The kinematical cuts are explained in the text.
} 
\label{fig2}
\end{figure}

\section{The measurement of the gluon density}
Dijet production in DIS at HERA  allows for a direct
measurement of the gluon density in the proton (via  $\gamma g \rightarrow
q\bar{q}$). Some investigations of
the feasibility of the parton density determination
from dijet production have been discussed in Ref.~[6]
(see also Ref.~[5]). 
Repond showed first results \cite{repond} 
of the gluon distribution determination
in NLO in the cone scheme.

\section{Forward Jet Production in the Low $x$ Regime}
Deep inelastic scattering with a measured forward jet 
with relatively large momentum fraction $x_{jet}$
(in the proton direction) and $p_T^{2\,lab}(j)\approx Q^2$
is expected to provide 
sensitive information about the BFKL dynamics at low $x$
\cite{mueller,allen1}.
In this region there is not much phase space 
for DGLAP evolution with transverse momentum ordering,
whereas large effects are expected for BFKL evolution in $x$.
In particular, BFKL evolution is expected to substantially enhance cross
sections in the region $x<<x_{jet}$ \cite{mueller,allen1}.
In order to extract information on the $\ln(1/x)$
BFKL evolution, one needs to show that cross section results based on fixed 
order QCD with DGLAP evolution are not sufficient to describe the data. 
Clearly, next-to-leading order QCD corrections to the DGLAP predictions
are needed to make this comparison between experiment and theory.

In Table~\ref{table2} we show numerical results for the multi jet cross 
sections with (or without) a forward jet. 
The LO (NLO) results are based on 
the LO (NLO) parton distributions from GRV \cite{grv} together with
the one-loop (two-loop) formula 
for the strong coupling constant.
Kinematical cuts are imposed to closely model the H1 event 
selection. More specifically, we require 
$Q^2>8~$GeV$^2$ , $x<0.004$,
$0.1 < y < 1$, an energy cut of $E(e^\prime)>11$~GeV on the scattered 
electron, and a cut on the pseudo-rapidity $\eta=-\ln\tan(\theta/2)$
of the scattered lepton of 
$ -2.868 < \eta(e^\prime)< -1.735$ 
(corresponding to $160^o < \theta(l^\prime) < 173.5^o$).
Jets are defined in the cone scheme (in the laboratory frame) with
$\Delta R = 1$ and $|\eta(j)|<3.5$.
We require a forward jet with $x_{jet}=p_z(j)/E_{P} > 0.05$,
$E(j)>25$ GeV, $0.5<p_T^2(j)/Q^2<4$,
and a cut on the pseudo-rapidity of 
$ 1.735< \eta(j)< 2.9$ 
(corresponding to $6.3^o < \theta(j) < 20^o$).
In addition all jets must have 
transverse momenta of at least  4 GeV in the lab frame
and 2 GeV in the Breit frame.
\begin{table}[t]
\caption{Cross sections for $n$-jet exclusive events
in DIS at HERA. See text for details.  
}\label{table2}
\vspace{2mm}
\begin{center}
\begin{tabular}{lcc}
        \hspace{0.8cm}
     &  \mbox{with  }
     &  \mbox{without  } \\
     &  \mbox{forward jet}
     &  \mbox{forward jet}\\
\hline\\[-3mm]
\mbox{1 jet (LO)}   & 0    pb &  9026 pb       \\
\mbox{2 jet (LO)}   & 19.3 pb &  2219 pb    \\
\mbox{2 jet (NLO)}  & 68   pb &  2604 pb    \\
\mbox{3 jet (LO)}   & 30.1   pb & 450  pb    \\
\end{tabular}
\end{center}
\end{table}

The cross sections of Table~\ref{table2} demonstrate first of all that
the requirement of a forward jet with large longitudinal momentum fraction
($x_{jet}>0.05$) and restricted transverse momentum ($0.5<p_T^2(j)/Q^2<4$)
severely restricts the available phase space, in particular for low jet
multiplicities. The 1-jet exclusive cross section vanishes at LO, due to the
contradicting $x<0.004$ and $x_{jet}>0.05$ requirements. For $x<<x_{jet}$,
a high invariant mass hadronic system must be produced by the photon-parton
collision and this condition translates into 
\begin{eqnarray}
2E(j)m_T\;e^{-y} &\approx& \hat{s}_{\gamma,parton} 
                 \approx Q^2\left({x_{jet}\over x}-1\right) >> Q^2\; ,
\end{eqnarray}
where $m_T$ and $y$ are the transverse mass and rapidity of the 
partonic recoil system, respectively. Thus a recoil system with substantial
transverse momentum and/or invariant mass must be produced and this 
condition favors recoil systems composed out of at least two additional 
energetic partons. 

As a result 
one finds very large fixed order perturbative QCD corrections (compare
2 jet LO and NLO results with a forward jet in Table~\ref{table2}). 
In addition, the LO $({\cal{O}}(\alpha_s^2))$ 3-jet cross section 
is larger than the LO $({\cal{O}}(\alpha_s))$
2-jet cross section.
Thus, the forward jet  cross sections in Table~\ref{table2} are dominated
by the $({\cal{O}}(\alpha_s^2))$ matrix elements.
The effects
of BFKL evolution must be seen and isolated on top of these fixed order QCD
effects. We will analyze these effects in a subsequent publication.


\end{document}